\documentclass[twocolumn,times]{aastex6}

\usepackage{natbib}
\usepackage{placeins}
\usepackage{amsmath}
\usepackage{mathtools}
\newcommand\degree{{^\circ}}
\newcommand\kms{km~s$^{-1}$}
\defcitealias{vogt1995}{V95}
\defcitealias{koch2007b}{K07b}
\defcitealias{kirby2011}{K11}

\begin{document}

\title{A Multi-Epoch Kinematic Study of the Remote Dwarf Spheroidal Galaxy Leo II$^{*}$}\thanks{*Observations reported here were obtained at the MMT Observatory, a joint facility of the University of Arizona and the Smithsonian Institution.}
\shorttitle{Multi-Epoch Kinematics of Leo II}
\shortauthors{Spencer et al.}
\author{Meghin E. Spencer\altaffilmark{1}$^,$\altaffilmark{2}, Mario Mateo\altaffilmark{1}, 
		Matthew G. Walker\altaffilmark{3}, Edward W. Olszewski\altaffilmark{4}}
\altaffiltext{1}{Department of Astronomy, University of Michigan, Ann Arbor, MI}
\altaffiltext{2}{Correspondence should be addressed to meghins@umich.edu}
\altaffiltext{3}{McWilliams Center for Cosmology, Department of Physics, Carnegie Mellon University, Pittsburgh, PA}
\altaffiltext{4}{Steward Observatory, The University of Arizona, Tucson, AZ}

\begin{abstract}
We conducted a large spectroscopic survey of 336 red giants in the direction of the Leo II dwarf galaxy using Hectochelle on the MMT, and conclude that 175 of them are members based on their radial velocities and surface gravities. Of this set, 40 stars have never before been spectroscopically observed. The systemic velocity of the dwarf is $78.3\pm0.6$ \kms{} with a velocity dispersion of $7.4\pm0.4$ \kms{} . We identify one star beyond the tidal radius of Leo II but find no signatures of uniform rotation, kinematic asymmetries, or streams. The stars show a strong metallicity gradient of $-1.53\pm0.10$ dex kpc$^{-1}$ and have a mean metallicity of $-1.70\pm0.02$ dex. There is also evidence of two different chemodynamic populations, but the signal is weak. A larger sample of stars would be necessary to verify this feature.
\end{abstract}


\section{Introduction}

Dwarf spheroidals (dSphs) are the smallest stellar systems that contain significant amounts of dark matter. They span a wide range of properties from the more luminous `classical' dSphs to the recently discovered `ultra-faint' dSphs, with half-light radii between $\sim30<r_{h}<700$ pc, total mass between $\sim2\times10^{5}<M_{\odot}<1\times10^{8}$, and luminosities between $\sim1\times10^{2}<L_{\odot}<1\times10^{7}$ \citep{mateo1998,mcconnachie2012}. This corresponds to mass-to-light ratios ranging from as little as 5 for some classical dSphs to as much as 5000 for the ultra-faints. These tiny systems are interesting not only for their physical diversity but also because they could be local analogs of the building blocks for larger galaxies, and can thus be used to explore the early evolution of galaxies.

The research collaboration represented in this paper has already analyzed many Milky Way dSphs (i.e. Leo I \citep{mateo2008}, Carina, Fornax, Sculptor, Sextans \citep{walker2009c}, and Draco \citep{walker2015}). In this paper we focus on Leo II, a classical dSph located far from the center of the Galaxy at 233$\pm$15 kpc \citep{bellazzini2005}. Most dSphs are found near the Milky Way (MW), but Leo II instead occupies a region of space that is dominated by star-forming dwarf irregular galaxies \citep[see, for example,][]{mateo1998}. Due to its large distance, it has often been questioned whether or not Leo II is gravitationally bound to the Milky Way \citep{demers1983}. Based off of its radial velocity and dSph morphology, it is reasonable to consider Leo II a MW satellite \citep[][their Figure 2]{mcconnachie2012}, but when taking into account the small galactocentric radial velocity component \citep{lepine2011,piatek2016} and lack of evidence for tidal disruption \citep{koch2007b}, it seems possible that Leo II has evolved in isolation with in the Local Group and is nearing the MW for the first time. In either scenario, Leo II is an interesting case to study. 

Since its discovery \citep{harrington1950}, Leo II has been the focus of many photometric studies. What started as only a few dozen individually detectable stars has evolved into massive studies of thousands of stars \citep[see, for example][]{bellazzini2005,komiyama2007,coleman2007,gullieuszik2008}. From these massive space-based and ground-based surveys, it has been concluded that Leo II has undergone little to no star formation in the last $\sim$7 Gyr \citep{mighell1996}; red clump stars are more centrally concentrated than blue horizontal branch stars \citep{bellazzini2005}; a mixture of stellar populations exists in the galaxy center while an older, more homogeneous population exists at larger radii \citep{komiyama2007}; and some minor isophotal twisting is present but there is no dynamical evidence for tidal distortion \citep{coleman2007}

Due to the relatively large distance of Leo II from the MW, far fewer stars have been observed spectroscopically. The first velocity measurements of only two very luminous red giant stars were published by \citet{suntzeff1986}, and shortly after came a study with five carbon stars \citep{zaritsky1989}. A more extensive study was carried out by \citet[][hereafter V95]{vogt1995}, which included 31 red giant branch members. Based on this dataset they concluded the bulk radial motion of the dwarf to be 76.0$\pm$1.3 \kms{} and the velocity dispersion to be 6.7$\pm$1.1 \kms{}. Furthermore they noted that the mass to light ratio in the V-band was 11.1$\pm$3.8, suggesting that the galaxy was embedded in a massive dark matter halo with mass of 9$\times$10$^{6}M_{\odot}$, similar to the known halo masses of other dwarfs \citep{mateo1993}, and consistent with more recent findings that the smallest dark matter halos are similar in mass \citep{walker2009a,strigari2008}. Since then, \citet[][hereafter K07b]{koch2007b} expanded upon the kinematic data of Leo II, observing 200 stars and concluding 171 of them were members. The precision of the individual velocity measurements was worse than \citetalias{vogt1995} by about 1 \kms{}, but with over five times more stars, they improved the precision of the systemic velocity measurement to $79.1\pm0.6$ \kms{} and the dispersion to $6.6\pm0.7$ \kms{}. They found no velocity gradient, velocity asymmetry, or signs of rotation, and therefore concluded that the galaxy has not been affected by tides. \citet{bosler2007} obtained low resolution spectra of 74 Leo II stars for the purpose of better understanding the chemical composition but lacked the necessary precision to report velocity measurements to better than $\sim50$ \kms{}. More recently, \citet{kirby2010} targeted 394 red giant branch stars in the direction of Leo II and determined 258 of them were members based on radial velocities. In a followup paper \citep[][hereafter K11]{kirby2011}, they focused on chemical abundances and notably derived a metallicity gradient of $-4.26\pm0.31$ dex deg$^{-1}$ in Leo II, which stood in contrast to the negligible slope found by \citet{koch2007a} for 52 stars.

In this paper we present new spectroscopic data with high precision from a large sample of red giant branch stars in Leo II. Details of our observing strategy, data reduction, and velocity extraction methods are found in Section \ref{observations}. Section \ref{kin_features} provides a kinematic analysis of the stars while Section \ref{chem_features} focuses on the chemistry of the stars. Section \ref{conclusions} contains concluding remarks and a summary of our findings.

\section{Observations and Data Processing} \label{observations}
\subsection{Photometry}

We used the 90prime imager \citep{williams2004} on the 2.3-meter Bok telescope at Steward Observatory in Arizona to collect photometry of Leo II. Stars were observed in the Washington $M$ and $I$ filters during February 2006. Data were processed in the usual way: subtracting an average bias frame, dividing by a normalized twilight flat-field, and adding repeated observations to remove cosmic rays.

\begin{figure}
\epsscale{1.1}
\plotone{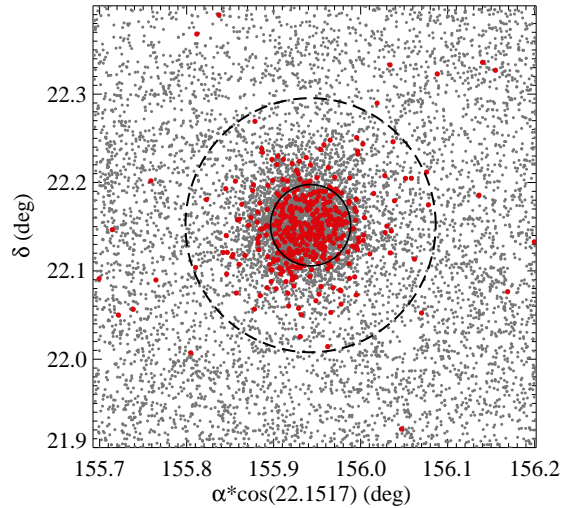}
\caption{Right ascension vs. declination of the targets observed photometrically. Stars highlighted in red were selected for spectroscopic followup based on the CMD in Figure \ref{cmd}. The core and tidal radii from \citet{komiyama2007} are shown as black solid and dashed lines, respectively.}
\label{photo_targets}
\end{figure}

\begin{figure}
\epsscale{1.1}
\plotone{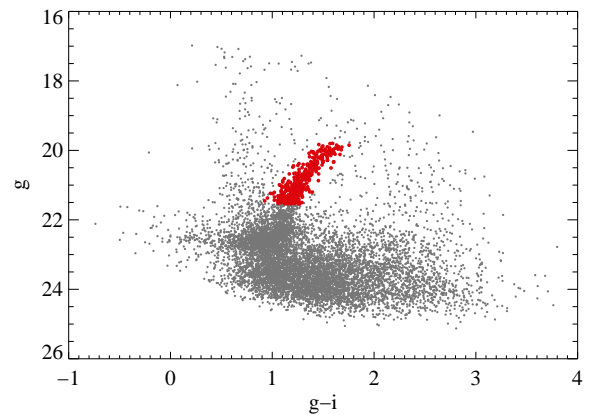}
\caption{Color magnitude diagram of stars in the direction of Leo II. The filters are Sloan g and i. Giant stars selected for spectroscopic followup are shown in red.}
\label{cmd}
\end{figure}

After processing, we used the DoPHOT software \citep{schechter1993} to get positions and magnitudes for objects in the images. The algorithm works by first fitting a user-supplied guess of the FWHM to all bright sources in the frame. After finding most of the obvious stars, it recalculates the FWHM and recomputes the brightness of each target. The magnitudes are recorded into a file along with the pixel coordinates and a set of $\chi^{2}$ values for an assumed single-star, multiple-star, or galaxy profile. 

We calibrated these instrumental $M$ and $I$ magnitudes by transposing them to SDSS apparent $g$ and $i$ magnitudes. Approximately half of the stars in our sample were listed in SDSS, so we used those stars to fit a three-term function relating the SDSS magnitudes, our magnitudes, and a color term. The best fitting transformations were $i=I+0.88(M-I)+7.52$ and $g=M-1.11(M-I)+8.52$. Table \ref{table_phot} lists the celestial coordinates and these apparent magnitudes. Figure \ref{photo_targets} shows all of the stars on the sky that we measured apparent magnitudes for. The red points are stars that we targeted for spectroscopic followup. They were selected on the basis of having $g$ magnitudes brighter than 21.55 and being confined within the red giant branch of the color magnitude diagram in Figure \ref{cmd}.

\begin{deluxetable}{c c c c}
\centering
\tablecaption{Photometric properties of stars in the direction of Leo II\label{table_phot}}
\tabletypesize{\scriptsize}
\tablehead{\colhead{$\alpha$} & \colhead{$\delta$} & \colhead{g} & \colhead{i} \\
\colhead{(J2000)} & \colhead{(J2000)} & \colhead{[mag]} & \colhead{[mag]} }
\startdata
168.3457500 &  22.1563333 & 18.20$\pm$0.03 & 19.80$\pm$0.03 \\ 
168.3078750 &  22.1473056 & 18.24$\pm$0.03 & 19.81$\pm$0.03 \\ 
168.3572083 &  22.0255278 & 18.10$\pm$0.05 & 19.85$\pm$0.04 \\ 
168.2647500 &  22.1929444 & 18.22$\pm$0.03 & 19.85$\pm$0.03 \\ 
168.5162500 &  22.1748056 & 18.37$\pm$0.03 & 19.83$\pm$0.03 \\ 
168.3427500 &  22.0601944 & 18.38$\pm$0.03 & 19.87$\pm$0.03 \\ 
168.3698333 &  22.2199722 & 18.24$\pm$0.04 & 19.89$\pm$0.03 \\ 
168.3795833 &  22.1084167 & 18.22$\pm$0.04 & 19.91$\pm$0.03 \\ 
168.3561667 &  22.2239722 & 18.36$\pm$0.03 & 19.90$\pm$0.03 \\ 
168.3367500 &  22.1429444 & 18.42$\pm$0.03 & 19.91$\pm$0.03 \\ 
\enddata
\tablecomments{This table is published in its entirety in the machine-readable format. A portion is shown here for guidance regarding its form and content.}
\end{deluxetable}

\subsection{Spectroscopy}

Spectroscopic observations were obtained with the Multiple Mirror Telescope (MMT) using Hectochelle, a multi-fiber, single-order echelle spectrograph \citep{szentgyorgyi1998}. The instrument can target up to 244 objects within a 1 degree field, and has an operational spectral range between 3,800 and 9,000 \AA. We used the RV31 filter for our observations, which isolates the spectral region spanning between $\sim$5150 and 5300 \AA{} and contains the MgI and Mg b features. Spectra were taken on five different runs between 2006 and 2013. Table \ref{table_obs} summarizes the observed fields, dates of observation, heliocentric julian dates, exposure times, and number of exposures.

Processing of the raw images was done with IRAF. The steps are identical to those in \citet{mateo2008}, but are briefly repeated here. The overscan region was subtracted from all images, and then trimmed out. Hectochelle has two amplifiers for each of its two CCDs, so data from the amplifiers for both CCDs was combined. Multiple exposures for each pointing were also combined to form a single, deeper image for each pointing, as listed in Table \ref{table_obs}. In doing so, cosmic rays could be simultaneously removed by a sigma clipping algorithm.

The fibers at the focal surface of the spectrograph collimator are staggered to allow for tighter packing, and thus the spectra need to be extracted before further reductions. Locations of the individual spectra on the CCD were traced by quartz lamp spectra that were taken after each science exposure. The quartz traces were allowed to shift en masse to align with the data. These shifted traces were used to extract science and calibration spectra. A fifth-order polynomial was used to produce a wavelength solution based off of 30-40 ThAr emission lines. Relative fiber throughputs were determined from twilight observations, as fibers were not evenly illuminated by the quartz lamp. The throughputs were then divided out. Lastly, sky spectra were recorded by unassigned fibers and combined to produce a master sky spectrum for each pointing, which was then subtracted from the science spectra. There was anywhere from 40 to 60 sky spectra in each pointing. This resulted in a set of 1,921 wavelength-calibrated, one-dimensional spectra with a resolution of 0.1 \AA/pix ($R\sim$25,000).

\begin{deluxetable*}{c c c c c c c}
\tablecaption{Hectochelle Observations of Leo II Fields\label{table_obs}}
\tabletypesize{\scriptsize}
\tablehead{\colhead{Field} & \colhead{$\alpha_{J2000}$\tablenotemark{a}} & \colhead{$\delta_{J2000}$\tablenotemark{a}} & \colhead{UT Date} & \colhead{HJD\tablenotemark{b}} & \colhead{N$_{exp}$\tablenotemark{c}} & \colhead{Exp. Time\tablenotemark{d}}  \\
 & \colhead{(hh:mm:ss.ss)} & \colhead{(dd:mm:ss.ss)} & \colhead{(dd/mm/yyyy)} & \colhead{(days)} & & \colhead{(seconds)} }
\startdata
LeoII-01 & 11:13:25.41 & +22:08:57.60 & 25/04/2006 & 2453850.67 &        3 &     8100 \\
LeoII-02 & 11:13:25.41 & +22:08:57.61 & 22/04/2007 & 2454212.79 &        2 &     5400 \\
LeoII-03 & 11:13:25.84 & +22:08:33.61 & 26/02/2008 & 2454522.72 &        3 &     7200 \\
LeoII-04 & 11:13:23.68 & +22:08:03.61 & 01/03/2008 & 2454526.82 &        3 &     7200 \\
LeoII-05 & 11:13:32.61 & +22:10:42.62 & 30/01/2011 & 2455591.92 &        5 &    12000 \\
LeoII-06 & 11:13:29.57 & +22:04:06.84 & 05/02/2011 & 2455597.96 &        4 &     9600 \\
LeoII-07 & 11:13:32.29 & +22:10:48.62 & 07/02/2011 & 2455599.81 &        2 &     4800 \\
LeoII-08 & 11:13:25.74 & +22:08:39.12 & 17/02/2013 & 2456340.93 &        4 &     2700 \\
LeoII-09 & 11:13:03.47 & +22:05:57.38 & 18/02/2013 & 2456341.94 &        3 &     2700 \\
\enddata
\tablenotetext{a}{central coordinates of field}
\tablenotetext{b}{at beginning of first sub-exposure}
\tablenotetext{c}{number of sub-exposures}
\tablenotetext{d}{exposure time summed over all sub-exposures}
\end{deluxetable*}

\subsection{Velocity Measures}

Most of our past papers analyzing Hectochelle data used fxcor --- a Fourier cross-correlation routine in IRAF --- to generate velocities from these spectra.  We have subsequently begun to use a new approach \citep{walker2015} that fits a library of synthetic spectra in order to estimate velocities as well as effective temperatures, surface gravities and metallicities. Since part of our analysis requires long baseline observations, we want to be certain that there are no systematic velocity differences between methods.  Therefore, we carried out our velocity measurements with both procedures to compare results quantitatively.  

\begin{figure}[b]
\epsscale{1.1}
\plotone{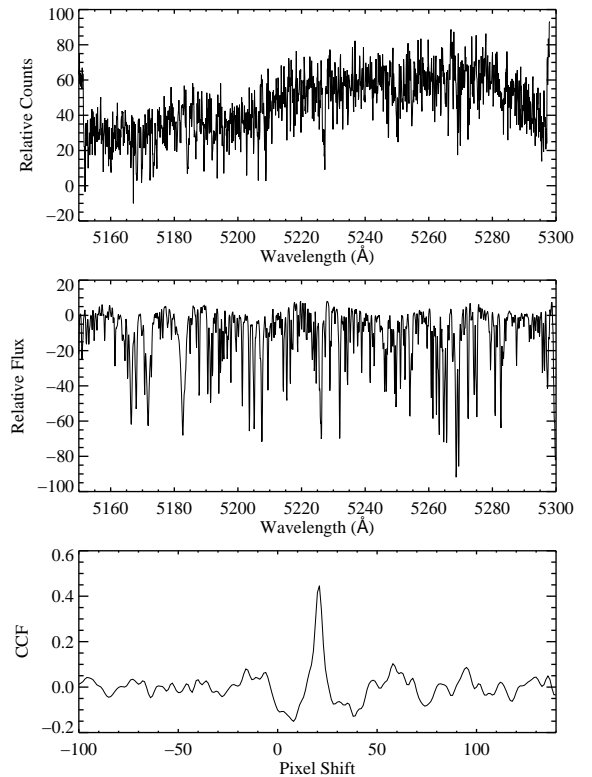}
\caption{Top: sample spectrum of one of our science targets. Middle: template spectrum used in fxcor, an IRAF task that performs a Fourier cross-correlation between a science spectrum and a template spectrum to determine a radial velocity. Bottom: Cross correlation function between the two spectra. The peak corresponds to the best shift between the two spectra and indicates the radial velocity for the science target.}
\label{fxcor}
\end{figure}

The fxcor analysis requires a suitable template spectrum to define the velocity zero point. The template that we used consists of co-added spectra acquired for various radial velocity standards with Hectochelle and is the same template used by \citet{mateo2008}; the co-added spectrum has S/N $>$ 350. Figure \ref{fxcor} illustrates the input and output of fxcor. The top and middle panels of Figure \ref{fxcor} show a sample science spectrum and the template spectrum, respectively. The bottom panel shows the cross correlation function, where the pixel shift at the highest peak corresponds to the redshift of the spectrum. The pixel shift is converted to a shift in wavelength, and thus a radial velocity. 

We refer the reader to \citet{walker2015} for a complete description of our newer method. Briefly, we obtain simultaneous estimates of radial velocity, effective temperature, surface gravity, and metallicity by fitting a library of smoothed, synthetic stellar spectra to each Hectochelle spectrum in pixel space. Following \cite{walker2015}, we use the library provided by \cite{lee2008a,lee2008b}, which was used to estimate stellar parameters for the SEGUE. The library is computed over a regular grid of $T_{\rm eff}$, $\log g$ and [Fe/H], and assumes a piecewise-linear relationship between [Fe/H] and [$\alpha$/Fe]. We use the software package MultiNest \citep{feroz2008,feroz2009} to sample parameter space and to sample the posterior probability distribution function (PDF) of our 15-dimensional model. For each parameter, we summarize the marginalized 1D PDF by recording the mean, variance, skewness and kurtosis. Following \citet{walker2015}, we use the skewness and kurtosis of the velocity distribution to reject poor-quality observations (see Section \ref{quality_control}).  

Our Leo II targets were each observed between one and seven times, giving us multiple measurements per star, often over many epochs. In total, we observed 727 spectra for 336 stars in the direction of Leo II.

\begin{figure}
\epsscale{1.1}
\plotone{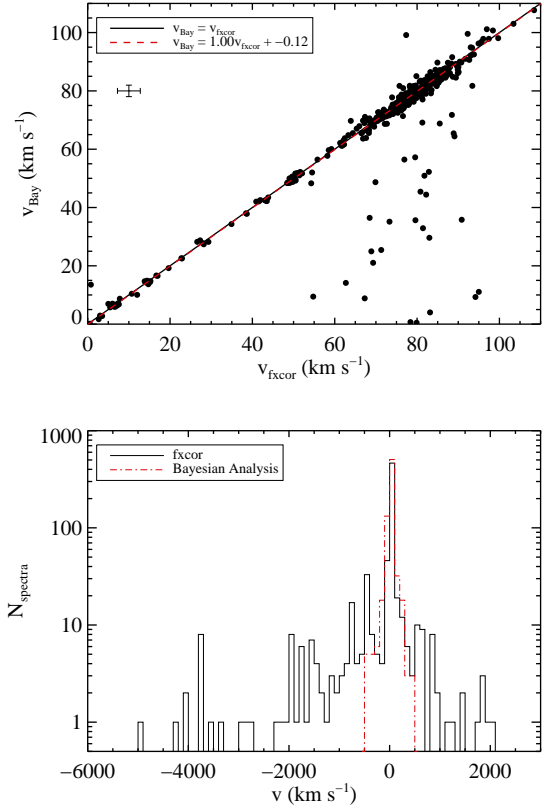}
\caption{Top: velocity measures extracted via Bayesian analysis plotted against velocity measures extracted by fxcor. Each dot represents a measure from a spectrum, so there may be multiple points per star. The median velocity error is 2.77 \kms{} for fxcor 1.97 \kms{} for Bayesian analysis; this is represented by the symbol in the top left. The black solid line marks the one-to-one line where the two measurements agree perfectly; the red, dashed line traces the best fit to the data when the slope has been set equal to 1. Bottom: number of spectra with a given velocity measure from fxcor (black, solid line) and Bayesian analysis (red, dash-dotted line).}
\label{fxcormcmc}
\end{figure}

We compare the velocity results from one method to the other in Figure \ref{fxcormcmc}. Error bars are not shown to increase plot readability; the median error for the fxcor method is 2.8 \kms{}, and for the Bayesian analysis method is 2.0 \kms{}. We fit a line with slope equal to unity and identify a very slight systematic offset of 0.13 \kms{}. This offset is well within the combined errors so we have chosen to apply no corrections to either set of velocity measurements. We choose to use the Bayesian approach for all further analysis because it extracts stellar atmospheric parameters and also has a more straightforward and self-consistent estimate of the errors.

\subsection{Quality Control}\label{quality_control}

\begin{figure}[b]
\epsscale{1.1}
\plotone{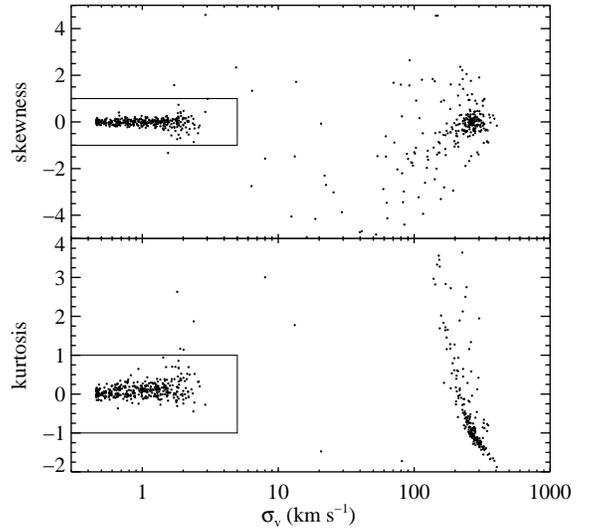}
\caption{Third and fourth moments are plotted against second moment of the posterior probability distribution functions from Bayesian analysis of the spectra. Measurements outside the black boxes are discarded as low quality and are not used for further analysis. The boundaries for the quality cuts are adopted from \citet{walker2015}.}
\label{kurtosis}
\end{figure}

Since we are ultimately interested in recovering the velocity dispersion of Leo II and to measure the velocity variability of its stars, we must be particularly careful in identifying and excising low-quality data from the sample. Shown in Figure \ref{kurtosis} are the velocity errors plotted against the skewness and kurtosis of the error distribution returned in the Bayesian analysis. In each panel, the points cluster in two groups, with good measures occupying the left side of the plot where error distributions are relatively narrow and Gaussian. For consistency, we adopt the same quality criteria as \citet{walker2015}. Thus, measurements used in the analysis of this paper have $\sigma_{v}<5$ \kms{}, $-1.0<\mathrm{skewness}_{v}<1.0$, and $-1.0<\mathrm{kurtosis}_{v}<1.0$. Of the 336 stars observed with MMT, 222 had velocity measures that met these criteria.

With the remaining velocity measurements, we combined any observations taken over multiple epochs to arrive at one average velocity per star, which is useful for determining membership. Velocities were weighted by the inverse of their variances and are expressed as $v = \sum \frac{v_{i}}{\sigma_{i}^{2}} / \sum \frac{1}{\sigma_{i}^{2}}$. Similarly, the error measurements were combined such that $\sigma=(\sum \frac{1}{\sigma_{i}^{2}})^{-1/2}$. Other spectral quantities and their errors were averaged in the same way, including [Fe/H], log(g), and T$_{eff}$. Up to seven different epochs of observations contributed to these average measurements. The averages are reported in Table \ref{data_table} and individual measures that went into these averages are listed immediately below the corresponding average. The columns are as follows: (1-2) celestial coordinates, (3) HJD, (4) heliocentric radial velocity and error, (5) effective temperature and error, (6) surface gravity and error, (7) metallicity and error, (8) the number of observations that went into the calculation of the average measurements, and (9) the star's membership status (see Section \ref{members}).

\begin{deluxetable*}{r r c r l r r c c}
\centering
\tablecaption{Spectroscopic properties of stars in the direction of Leo II\label{data_table}}
\tabletypesize{\scriptsize}
\tablehead{\colhead{$\alpha$} & \colhead{$\delta$} & \colhead{HJD} & \colhead{$v$} & \colhead{T$_{eff}$} & \colhead{log(g)} & \colhead{[Fe/H]} & \colhead{N$_{obs}$} &\colhead{member?} \\
\colhead{(J2000)} & \colhead{(J2000)} & \colhead{(-2450000 days)} & \colhead{[\kms]} & \colhead{[K]} & \colhead{[dex]} & \colhead{[dex]} & & }
\startdata
168.228745 & 22.368030 & 5059.35 & 208.67  $\pm$ 0.87 &         5209 $\pm$          223 & 3.15 $\pm$ 0.43 & -1.78 $\pm$ 0.26 &       2 & N \\
 & & 4526.81 & 208.87  $\pm$ 1.00 &         5153 $\pm$          238 & 3.31 $\pm$ 0.46 & -1.94 $\pm$ 0.29 &  &  \\
 & & 5591.90 & 208.07  $\pm$ 1.77 &         5621 $\pm$          644 & 2.19 $\pm$ 1.13 & -1.58 $\pm$ 0.64 &  &  \\
168.332825 & 22.139639 & 4526.81 & 71.08   $\pm$ 2.19 &         6347 $\pm$          601 & 1.46 $\pm$ 0.68 & 0.17  $\pm$ 0.53 &       1 & Y \\
168.361542 & 22.140249 & 4526.81 & 76.11   $\pm$ 2.64 &         5636 $\pm$          438 & 4.65 $\pm$ 0.75 & -0.73 $\pm$ 0.51 &       1 & N \\
168.331375 & 22.146915 & 5591.90 & 75.00   $\pm$ 0.91 &         4800 $\pm$          167 & 0.98 $\pm$ 0.27 & -1.26 $\pm$ 0.22 &       1 & Y \\
168.392960 & 22.153250 & 5591.90 & 71.64   $\pm$ 0.78 &         4622 $\pm$          147 & 0.92 $\pm$ 0.24 & -2.11 $\pm$ 0.18 &       1 & Y \\
168.353699 & 22.122332 & 5969.43 & 78.11   $\pm$ 0.90 &         4946 $\pm$          173 & 1.02 $\pm$ 0.23 & -0.96 $\pm$ 0.23 &       2 & Y \\
 & & 5597.94 & 79.17   $\pm$ 2.17 &         5492 $\pm$          714 & 2.81 $\pm$ 0.92 & -2.43 $\pm$ 0.73 &  &  \\
 & & 6340.91 & 77.89   $\pm$ 0.99 &         4912 $\pm$          179 & 0.90 $\pm$ 0.23 & -2.22 $\pm$ 0.24 &  &  \\
\enddata
\tablecomments{This table is published in its entirety in the machine-readable format. A portion is shown here for guidance regarding its form and content.}
\end{deluxetable*}

\section{Kinematic and Chemical Analysis} \label{analysis}
\subsection{Defining Membership} \label{members}

\begin{figure}
\epsscale{1.1}
\plotone{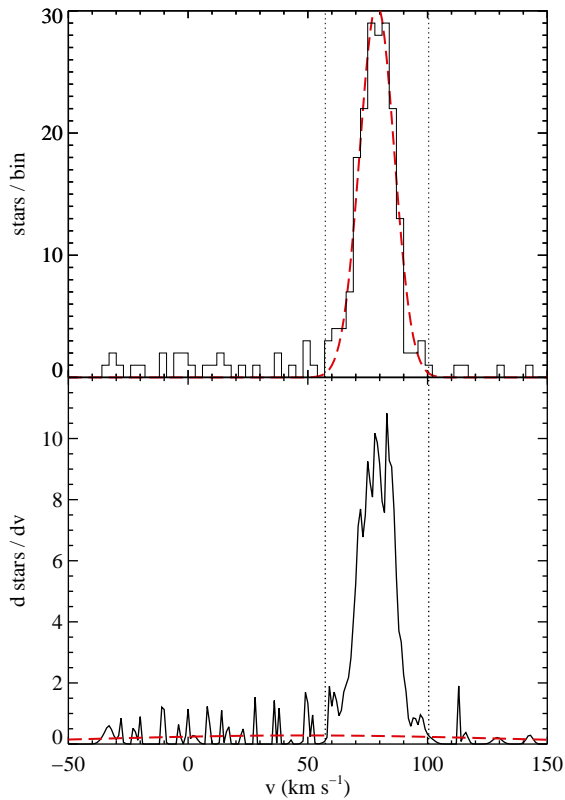}
\caption{Top: the number of stars in each velocity bin are shown here in solid black. The Gaussian fit to the histogram is plotted as a red dashed line. The center and standard deviation of the Gaussian are 78.9 \kms{} and 7.2 \kms{} respectively. Stars within 3$\sigma$ of the center velocity are considered velocity members; this boundary is marked by the vertical dotted lines. Bottom: each star is represented by a gaussian distribution with unit area and the sum of these Gaussians is the solid black line. The red dashed line is a distribution of $\sim25,000$ stars from \cite{robin2003}, scaled such that the integrated area represented by stars with velocities between -50 and 50 km/s is equal to the area of the stars from this paper in the same range. The area under the red dashed curve between the Leo II velocity boundaries is 11.3, suggesting there should be 11 foreground stars.}
\label{hist}
\end{figure}

To separate stellar members from non-members we first employed a simple velocity cut. Figure \ref{hist} shows a histogram of the averaged velocity measures, so that there is one data point per star. We fit a three-parameter gaussian to the histogram of the form $f(v)=a_{0}\exp(-\frac{(v-a_{1})^{2}}{2a_{2}^{2}})$. The best fit parameters were $a_{0}$=30.6, $a_{1}$=78.9 \kms{}, and $a_{2}$=7.2 \kms{}. Stars with radial velocities that fall within 3$\sigma$ of the center (within the range $57.3 < v < 100.5$ \kms{}) were taken to be likely members of Leo II, while stars outside this range were assumed to be foreground Milky Way halo stars. This boundary is marked as two vertical dotted lines in Figure \ref{hist}. Employing this cut yielded 186 velocity members of Leo II.

There is expected to be a small number of apparent member stars that are actually halo stars with projected positions and velocities matching those of Leo II. We quantified this fraction by using the Besan\c{c}on models of the MW halo \citep{robin2003}. We produced a sample of 25,000 stars that would exist along the line of sight toward Leo II according to the model. We then computed a generalized histogram, whereby each star is represented by a gaussian curve with unit area, centered on the listed Besan\c{c}on velocity and having a standard deviation equal to the median of the weighted MMT velocity errors (0.94 \kms{}). The $\sim$25,000 gaussians were summed up to produce a single smooth distribution. A generalized histogram was also produced for our 222 stars with observed radial velocities, but using the velocity errors as the standard deviations. We normalized the Besan\c{c}on distribution by requiring the area under the curve between $-50<v<50$ \kms{} ---the typical velocity range of Milky Way foreground stars---to be equal to the area occupied by our observed stars within the same region. This normalized distribution of model MW halo stars is shown as a red dashed line in the lower panel of Figure \ref{hist}; the generalized gaussian histogram for our observed stars is marked by a black solid line. By integrating the red distribution over the velocity range of accepted Leo II membership, we estimated that there should be 11 halo stars with velocities and positions similar to those of Leo II. 

\begin{figure}
\epsscale{1.1}
\plotone{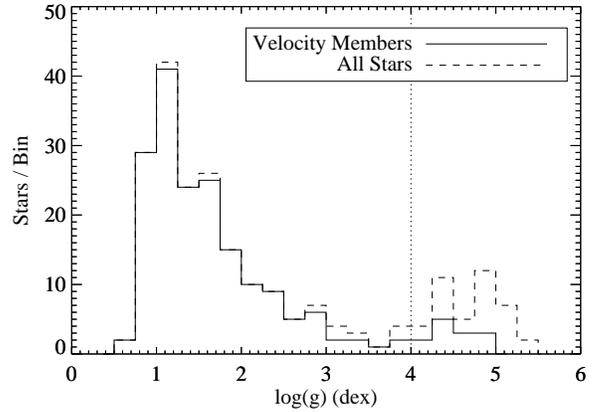}
\caption{The number of stars per log(g) bin of width 0.25 dex is plotted for all stars (dashed line) and for only velocity members (solid line). Most stars that are velocity non-members have high surface gravities, as expected for MW halo stars. For this reason we consider all stars with log(g)$>$4 non-members.}
\label{hist_logg}
\end{figure}

Due to this contamination, we choose to apply one more cut on the data based on stellar surface gravities.  As can be seen in Figure \ref{hist_logg}, the majority of stars that were flagged as non-members according to radial velocities also have high surface gravities. This is expected since the stars we targeted should be on the red giant branch if they are members of Leo II, but will be dwarfs if they are foreground Milky Way stars. Therefore, our final requirement to be considered a member is that stars have $\log(g)\leq4$. This removes 11 stars from the sample, which is equal to the expected contamination from the Besan\c{c}on model.

Utilizing these cuts in velocity and surface gravity ($57.3 < v < 100.5$ \kms{} and $\log(g)\leq4$), we have a total sample of 175 member stars. This is the set of stars that we will use for the kinematic and chemical analyses in this paper.

We compared our velocities with others published in \citetalias{vogt1995}, \citetalias{koch2007b}, and \citetalias{kirby2011} (obtained via private communication). There were 22, 94, and 94 stars that were observed in both the respective studies and ours. The offsets between our data and previous studies were 0.84 \kms{} for \citetalias{vogt1995}, 0.66 \kms{} for \citetalias{koch2007b}, and 0.61 \kms{} for \citetalias{kirby2011}.  All of these offsets are smaller than the median errors of the datasets, suggesting good agreement between studies. More details on this comparison can be found in \citet{spencer2016b}.

\newpage
\subsection{Kinematic Features}\label{kin_features}

The systemic velocity and velocity dispersion of Leo II was calculated following the method of maximum likelihood laid out by \cite{walker2006}. The observed quantities are found by maximizing the natural logarithm of the joint probability function of the two being drawn from gaussian distributions. Following the notation of \cite{walker2006},
\begin{equation}
\mathrm{ln}(p) = -\frac{1}{2}\sum_{i=1}^{N}\mathrm{ln}(\delta^{2}_{i}+\sigma^{2}_{p})-\frac{1}{2}\sum_{i=1}^{N}\frac{(v_{i}-\langle v\rangle)^2}{(\delta^{2}_{i}+\sigma^{2}_{p})}-\frac{N}{2}\mathrm{ln}(2\pi) .
\end{equation}
Here, $\langle v\rangle$ is the systemic velocity and $\sigma_{p}$ is the velocity dispersion. $v_{i}$ and $\delta_{i}$ are the radial velocity and corresponding error for star $i$ selected from a total of $N=175$ stars. 
Errors were calculated through a covariance matrix with the variances of $\langle v\rangle$ and $\sigma_{p}$ as the diagonal elements. Further details can be found in \cite{walker2006}. This yielded a systemic velocity of 78.5$\pm$0.6 \kms{} and a velocity dispersion of 7.4$\pm$0.4 \kms{} over the full tidal radius of the dwarf. Both of these measurements agree with the best fit values of the Gaussian in Section \ref{members} to within 1-$\sigma$. Our systemic velocity falls comfortably between those from \citetalias{vogt1995} and \citetalias{koch2007b}, which are 76.0$\pm$1.3 \kms{} and 79.1$\pm$0.6 \kms{}, respectively. The velocity dispersion is also consistent within 1-$\sigma$ of both \citetalias{vogt1995} (6.7$\pm$1.1 \kms{}) and \citetalias{koch2007b} (6.6$\pm$0.7 \kms{}). The weighted average between these three measures is 78.5$\pm$0.4 \kms{} for the systemic velocity and 7.2$\pm$0.3 \kms{} for the velocity dispersion.

\begin{figure}
\epsscale{1.1}
\plotone{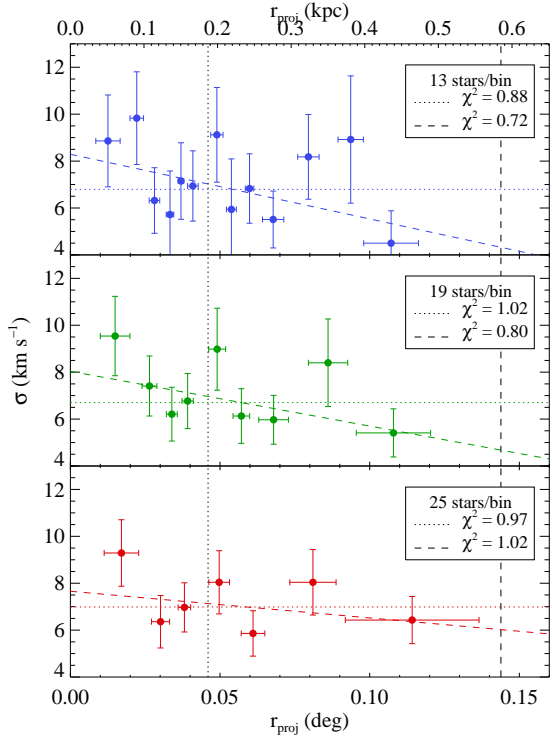}
\caption{The velocity dispersion profile is plotted using three different bin sizes: 12 stars per bin (top), 19 stars per bin (middle), 25 stars per bin (bottom). Errors in the radial direction are the standard deviations of the projected radii for stars in each bin; errors in the velocity dispersion come from the covariance matrix discussed in Section \ref{kin_features}. Black vertical dotted and dashed lines are the core and tidal radii respectively. We fit a flat (dotted) and sloped (dashed) line to each of the profiles. The reduced chi squared values of the fits are indicated in the plot legends.}
\label{vel_disp}
\end{figure}

We produced three radial velocity dispersion profiles using bin sizes of 13, 19 and 25 stars per bin. The dispersions for the stars within the bins were found by a similar method as described above except we set the systemic velocity equal to the value calculated using all Leo II member stars, which was 78.5 \kms{}. These profiles can be seen in Figure \ref{vel_disp}. The error bars in the radial direction are the standard deviations of the radii in those bins. Errors in the velocity dispersion were found using the same method described above with the covariance matrix. We fit a flat line and a sloped line to each of the velocity dispersion profiles, which are plotted as a dotted and dashed line respectively. The reduced $\chi^{2}$ between these lines and the data are listed in the top right corners of the plots. In all cases, the data are fit equally well by a flat line as a sloped line. Additionally, the error bars on the sloped lines are large enough such that the sloped lines are indistinguishable from a constant dispersion at the 1-$\sigma$ level. Therefore we conclude that the velocity dispersion remains flat at all radii regardless of bin size. These results are in good agreement with \citetalias{koch2007b} who also found a flat profile. 

The velocity dispersion can be used to produce a simple mass estimate for Leo II. We used the estimator in Equation 10 of \citet{walker2009c} which reduces to $M(r_{half})=2.5r_{half}\sigma^2/G$ when evaluated at the half-light radius. This method assumes that the stars are distributed as a Plummer sphere and have an isotropic velocity distribution with constant dispersion, all of which are reasonable for Leo II. We used $r_{half}=176\pm42$ pc \citep{mcconnachie2012} and found $M(r_{half})=5.6\pm1.4\times10^6M_\odot$. Dividing this mass estimate by half of the total luminosity \citep[$7.4\pm2.0\times10^{5}~L_{\odot}$, ][]{coleman2007} yields a mass to light ratio of $(M/L)_{V}=15.2\pm5.5$ in solar units, consistent with previous results.

\begin{figure}
\epsscale{1.1}
\plotone{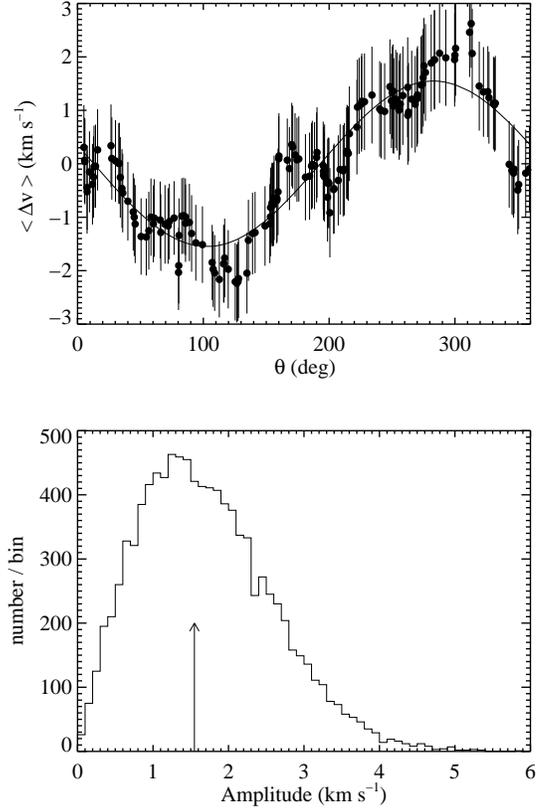}
\caption{Top: the difference between the average velocity of stars on either side of a bisecting line plotted against the position angle of that line. 0 degrees is North; 90 degrees is East. The vertical error bar is treated as the standard deviation of the stellar velocities divided by the square root of half the number of stars. The solid line is the best fit sinusoid to the trend and has an amplitude of 1.55 \kms{}. Bottom: we completed $10^{4}$ Monte Carlo realizations and performed the same rotation analysis on them. The amplitudes of these simulations is plotted as a histogram. The vertical arrow marks the location of the amplitude that we recovered for Leo II. Amplitudes larger than this are expected to be present in 52\% of non-rotating systems, therefore there is no statistically significant evidence for uniform rotation in Leo II. }
\label{bisect}
\end{figure}

With this sample we can test for signatures of ordered rotation within the dwarf. To do this, we sliced the dwarf in half and computed the difference between the average velocity for each of the two halves. The position angle, $\theta$, of the bisecting line was rotated through 360$\degree$, with 0$\degree$ marking the meridian through the center of Leo II. A sinusoidal pattern is distinguishable as seen in the top panel of Figure \ref{bisect}, and was fitted with $\langle v\rangle=a_{1}\sin(\theta + a_{2})$, where $a_{1}=1.55$ \kms{} (amplitude) and $a_{2}=167.1 \degree$ (phase).

To determine the likelihood that a 1.55 \kms{} signal could be produced by chance, we generated 10,000 Monte Carlo simulations with stellar positions equal to those of our observations and velocities drawn from a gaussian having standard deviation equal to the velocity dispersion of Leo II. The bottom panel of Figure \ref{bisect} is a histogram of the amplitudes from these simulations. 52\% of the trials have amplitudes larger than what we find in Leo II, thus the signal we find is only significant at a 0.64 $\sigma$ confidence level. From this we conclude that Leo II has no statistically significant, ordered rotation. \citetalias{koch2007b} recovered a slightly stronger signal with an amplitude of $\sim$2 \kms{} and a position angle at 16.5$\degree$. They ran similar Monte Carlo tests and found that 87\% of the tests had an amplitude greater than 2 \kms{} with a highly variable position angle for the peak velocity signal. Thus our conclusion matches that of \citetalias{koch2007b}.

We also completed a test to identify if any stars clumped in 3-D (ra, dec, radial velocity) phase space, as such features might indicate a more interesting merger history for Leo II \citep{coleman2004,assmann2013}. We considered the similarity of velocities between stars and their nearest neighbors. For each star we counted how many of its neighbors within a given radius had velocities similar to that star. We considered radii between 10 and 50 arcseconds and velocities within 0.5 to 2 \kms{} of the central star. To understand if the number of stars in each iteration was significant, we randomly reassigned the velocities to different spatial positions 10,000 times and completed the same exercise. In all cases, no signatures of clumping were found at statistically significant levels.

\begin{figure}
\epsscale{1.1}
\plotone{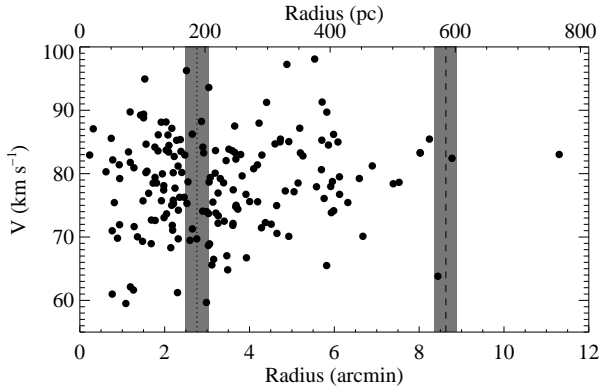}
\caption{Stellar radial velocity versus projected radius from the dSph center. The King core and tidal radii are plotted as vertical dotted and dashed lines respectively. Values for the radii are taken from \cite{komiyama2007}. Errors on these measurements are shown as gray shaded bars. Several stars are found near the tidal radius, but only one is located well beyond this limit. The upper x-axis scale in parsecs is based on a distance of 233 kpc \citep{bellazzini2005}.}
\label{vvsr}
\end{figure}

Having no rotation, clumps, or otherwise interesting kinematic substructure, the only remaining dynamic peculiarity that we find in Leo II is one star located beyond the tidal radius, as can be seen in Figure \ref{vvsr}. The separation of this star from the dwarf center is 1.3 times the tidal radius. A couple other stars are located near the tidal radius, but only one is positioned at least 3-$\sigma$ beyond that boundary. The velocity ($v$=83.0 \kms{}), surface gravity (log(g)=1.03 dex), and metallicity ([Fe/H]=-1.66 dex) of this star are all very close to the mean values for the dwarf. While photometric studies have found stellar overdensities beyond the tidal radius \citep{komiyama2007}, this is the first extratidal star with kinematic evidence supporting its membership. Our star and the 4-star photometric clump found in \citet{komiyama2007} are separated by $\sim5$ arcmin but are both located $\sim11$ arcmin from the center on the western side of the dwarf.

The existence of tidal stars in other dwarfs like Ursa Minor \citep{munoz2005} and Carina \citep{munoz2006a} have been attributed to tidal disruption from the Milky Way. The interpretation for the tidal star in Leo II is slightly different as the dwarf galaxy is located so far away; it is likely encountering the inner parts of the MW dark matter halo for the first time \citep{lepine2011} and would not yet exhibit such features. Recent evidence suggests that Leo II is falling into the Milky Way in a tidal stream of satellites comprised of Leo IV, Leo V, Crater, and Crater 2 \citep{torrealba2016}. The positions of both our star and the photometric clump are not aligned with the great circle that connects all five satellites, ruling out the notion that they were caused by streaming motion. Instead, it seems plausible that our star and the clump in \cite{komiyama2007} are remnants of the interactions between these satellites prior to their disruption. The best interpretation for the 4-star cluster in \cite{komiyama2007} was that a small globular cluster merged with Leo II, which would fit with this scenario. Future studies of Leo II may wish to obtain velocities for stars beyond the tidal radius for more conclusive evidence regarding the nature of these features.

\subsection{Chemical Features}\label{chem_features}

\begin{figure}
\epsscale{1.1}
\plotone{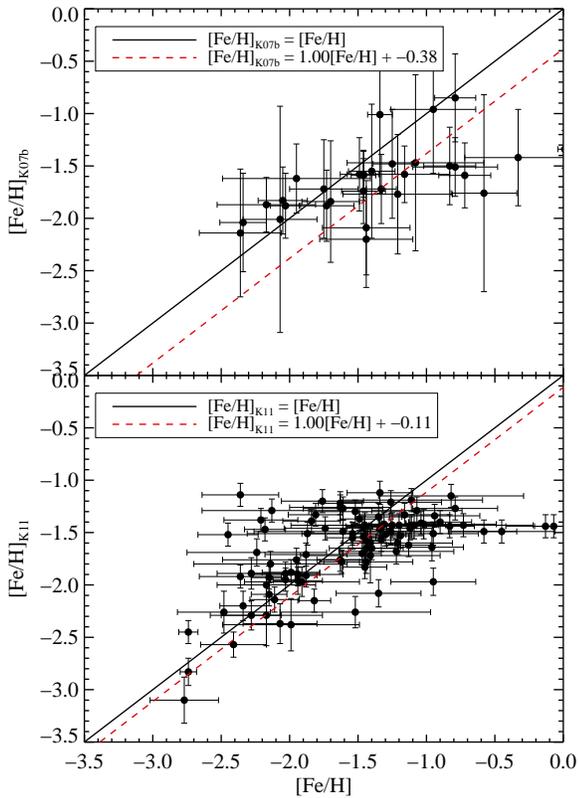}
\caption{A comparison between our [Fe/H] measures and those from \citetalias{koch2007b} (top panel) and \citetalias{kirby2011} (bottom panel).  Not all stars listed in \citetalias{koch2007b} have published metallicities. The black lines are where stars would fall if there were perfect agreement between the studies. The red lines are the best fits to the data with slope set equal to 1.}
\label{compare_feh}
\end{figure}

[Fe/H] stellar metallicities were reported in both \citetalias{koch2007b} and \citetalias{kirby2011}. Many of our stars also exist in those papers, so we completed a quick comparison to see if there were any major differences between them. In the top panel of Figure \ref{compare_feh}, we plot [Fe/H] from \citetalias{koch2007b} against our own data. The offset between them is 0.38 dex---as large as the mean scatter---so no real correlation between them can be identified. Since the spectral resolution of their study was less than ours, this result is not surprising. The comparison with \citetalias{kirby2011} is much better for low metallicities, with an offset of only 0.11 dex, but their distribution saturates at [Fe/H]$\approx$-1 while ours extends to higher metallicity.

We inspected our spectra of these high metallicity stars and found that the sky subtraction was poor, leading to some absorption features having negative flux. As a result, these metallicity measurements got pushed to larger values. We identify 13 such stars in our sample that have metallicities larger than -0.7 dex and remove them for the remaining chemical analysis. 

\begin{figure}
\epsscale{1.1}
\plotone{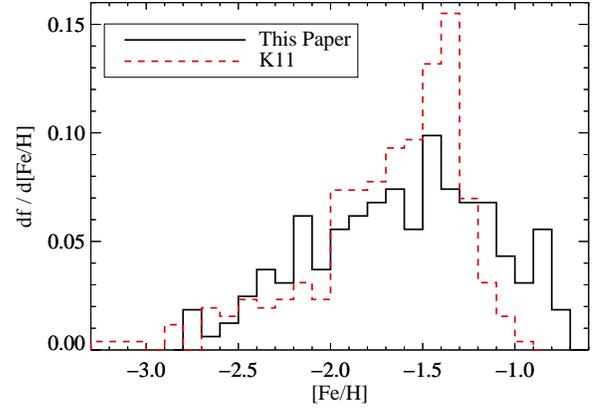}
\caption{Metallicity distribution function of Leo II members with [Fe/H] measurements in our dataset (black) and from \citetalias{kirby2011} (red dashed). The true number of stars per bin can be recovered by multiplying the values on the y-axis by the number of stars in the sample (162 for solid black or 258 for red dashed).}
\label{mdf}
\end{figure}

In Figure \ref{mdf} we plot the metallicity distribution function (MDF) for our data and a separate one for \citetalias{kirby2011} as a comparison. Both of these datasets have similar spatial distributions and thus we might expect the MDF to be comparable for our stars and those in \citetalias{kirby2011}. The mean metallicity weighted by the measurement uncertainties in our data is $\langle$[Fe/H]$\rangle=-1.70\pm 0.02$ dex. The standard deviation uncorrected for measurement errors is 0.48 dex. Correcting for measurements uncertainties as done by \citetalias{kirby2011} instead yields a width of 0.40 dex. The skewness of the distribution is -0.27, which indicates a low-metallicity tail. The (excess) kurtosis is -0.67, which means the MDF is less peaked than a normal distribution, which has a kurtosis of 0. Many of these MDF characterizations are discrepant from the ones published in \citetalias{kirby2011}, who report a mean metallicity of $\langle$[Fe/H]$\rangle=-1.62\pm 0.01$ dex. Their standard deviation and spread corrected for measurement uncertainties are 0.42 and 0.37 dex, and their skewness and kurtosis are -1.11 and 1.10 respectively, implying that our MDF is slightly wider, less peaked, and has a shorter low-metallicity tail than the MDF in \citetalias{kirby2011}. All of these features can be seen in Figure \ref{mdf}.

\begin{figure}
\epsscale{1.1}
\plotone{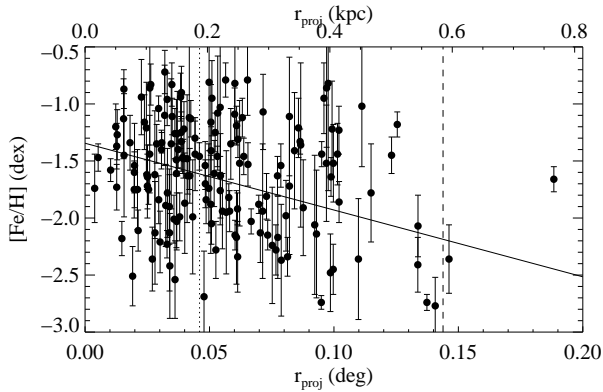}
\caption{The metallicities of the member stars are plotted against their separations from the dSph center. Core and tidal radii are shown as vertical dotted and dashed lines, respectively. The solid black line is a best fit to the data and has a slope (metallicity gradient) of $-5.85\pm0.39$ dex deg$^{-1}$, or $-1.53\pm 0.10$ dex kpc$^{-1}$ assuming a distance of 233 kpc \citep{bellazzini2005}.}
\label{fehvsr}
\end{figure}

Strong metallicity gradients have been found in about half the classical dSphs of the MW, with the slope in Leo II being the steepest \citepalias{kirby2011}. On the other hand, \citetalias{koch2007b} reports no such gradient for Leo II. In Figure \ref{fehvsr} we plot the metallicity versus radius of the stars in our sample, once again excluding the 13 stars with erroneous high-metallicity measurements. We fit a flat and sloped line to the data and find that neither provides a very good fit, with reduced $\chi^2$ values of 7.6 and 6.3 respectively. The metallicity gradient listed in \citetalias{kirby2011} was determined by the slope of a 2-parameter best fit line; even though our sloped line is a poor fit to the data, we continue with the analysis to provide a side by side comparison of our metallicities and those in \citetalias{kirby2011}. The best fitting sloped line to our data yielded a gradient of $-5.85\pm0.39$ dex deg$^{-1}$, or $-1.53\pm 0.10$ dex kpc$^{-1}$ using a distance of 233 kpc \citep{bellazzini2005}. This slope is somewhat steeper than the metallicity gradient published by \citetalias{kirby2011}, who found $-4.26\pm0.31$ dex deg$^{-1}$ ($-1.11\pm0.08$ dex kpc$^{-1}$, for a distance of 219 kpc). Regardless of the discrepancy between the slope measurements, both indicate that there is a large metallicity gradient with metal-rich stars clustering toward the center of the galaxy. The existence of a metallicity gradient agrees with the photometric findings that red clump stars are more centrally clustered than blue horizontal branch stars \citep{bellazzini2005}. This connection arises because high- and low-metallicity red giant branch stars are, respectively, the progenitors for red clump and blue horizontal branch stars.

\begin{figure}
\epsscale{1.1}
\plotone{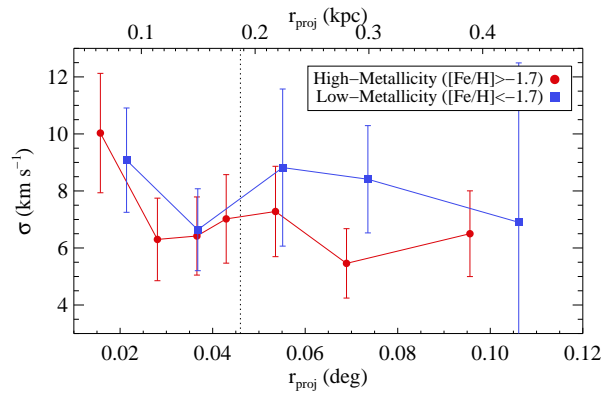}
\caption{The velocity dispersion profiles for high-metallicity stars (red circles, [Fe/H]~$>-1.70$) and low-metallicity stars (blue squares, [Fe/H]~$<-1.70$). There are 13 stars in each bin. The core radius is shown by a vertical dotted line.}
\label{veldisp_feh}
\end{figure}

The tendency for high-metallicity stars to be in a colder, less extended distribution than the low-metallicity stars is seen in many resolved dwarfs. For example, Fornax \citep{battaglia2006}, Sculptor \citep{battaglia2008}, and Sextans \citep{battaglia2011}. We explored the possibility that Leo II might also show different chemo-dynamic populations by first splitting the stars into two groups by the [Fe/H] mean, such that high-metallicity stars have [Fe/H] $> -1.7$ and low-metallicity stars have [Fe/H] $< -1.7$. Then we plotted the velocity dispersion profiles for these selections, which can be seen in Figure \ref{veldisp_feh}. The dispersion for the low-metallicity stars (blue points) is always larger than the dispersion for the high-metallicity stars (red points), but given the large error bars, the two profiles are consistent with being the same. We also calculated the overall dispersion for each of the two supposed populations. The high-metallicity stars have a velocity dispersion of $7.04\pm0.54$ \kms{} and the low-metallicity stars have a dispersion of $8.13\pm0.74$ \kms{}. Once again, we find that the values suggest the high-metallicity stars are kinematically colder, but when the errors are considered it is only a $1.2~\sigma$ detection.

As one final test, we allowed the high and low metallicity splitting value to vary from [Fe/H] $=-2.2$ dex to [Fe/H]$ =-1.1$ dex in steps of 0.05 dex, as opposed to fixing it at the mean of -1.7 dex. In all but one case, the high-metallicity dispersion was less than the low-metallicity dispersion, but the error bars made the results not significant beyond $1.6~\sigma$  at the most. Furthermore, we tried removing any stars with metallicities within 0.1 dex of the split value, as these stars might dampen the signal. The results were the same as before, though with slightly less significance. Taking all of the evidence together, it seems possible that there might be a slight chemo-dynamic bifurcation whereby high-metallicity stars have a larger velocity dispersion than low-metallicity stars due to the fact that this was repeatedly the trend in our data. Nevertheless, our large error bars caused by sample size and measurement error make it impossible to state this claim with more than $\sim1 \sigma$ confidence. A larger sample of stars with individual velocity precisions $\lesssim2$ \kms{} will be needed to explore the question of multiple chemical populations in Leo II definitively. 

\section{Summary and Conclusions} \label{conclusions}

In total we spectroscopically observed 336 stars within Leo II and determined that 175 of them are members based on radial velocities ($57.3<v_{mem}<100.5$ \kms{}) and surface gravities ($\log(g)_{mem}\leq4$). Many of the observed stars extend beyond the tidal radius of the dwarf into regions not previously studied by other publications, however only one of these extratidal stars is a member according to our constraints. Other than this one star, there are no signs of tidal disruption or rotation in the dwarf. By maximizing the likelihood that the velocities of these stars were drawn from a normal distribution, we determined that the systemic velocity of the dwarf is $78.5\pm0.6$ \kms{}, and its velocity dispersion is $7.4\pm0.4$ \kms{}. The velocity dispersion profile is consistent with being flat even when using three different bin sizes, suggesting that Leo II is embedded in a massive dark matter halo that extends well beyond the tidal radius. Using a simple King mass estimate, we determined the corresponding mass for Leo II to be $M(r_{half})=5.6\pm1.4\times10^6M_\odot$ and its mass to light ratio to be $(M/L)_{V}=15.2\pm5.5$  in solar units. 

The mean metallicity of the member stars is $\langle\mathrm{[Fe/H]}\rangle=-1.70\pm0.02$ dex, which is only slightly higher than average for dSphs of the MW. The shape of the metallicity distribution function is wider, less peaked, and has a shorter low-metallicity tail than the MDF reported in \citetalias{kirby2011}. Additionally, we found that Leo II has a strong metallicity gradient of $-5.85\pm0.39$ dex deg$^{-1}$ ($-1.53\pm0.10$ dex kpc$^{-1}$). Lastly we used three different tests to look for differences in the dynamics of high- and low-metallicity stars. In all cases, the results had low signal but were consistent with a model that has correlated metallicities and kinematics.

The dataset that we have compiled adds eight epochs of observation between the years 2006 and 2013 for stars in Leo II. 50 of these stars were observed on more than one occasion, with the maximum number of repeat observations being 5. Combining the data from \citetalias{vogt1995}, \citetalias{koch2007b}, \citetalias{kirby2011}, and this paper, there are 372 stars that are likely members of Leo II and 196 stars with repeat observations. Given the wealth of temporal radial velocity measurements, it is now possible to determine the binary fraction of stars in Leo II and evaluate the impact that these stars have on the measured velocity dispersion. This analysis will be carried out in a follow-up paper \citep{spencer2016b}.

\newpage
\acknowledgements

The authors would like to thank the MMT operators and staff for their help on observing runs between the years 2006-2013, along with Andy Szentgyorgyi, Dan Fabricant, Gabor Furesz, Nelson Caldwell, Perry Berlind, and Michael Calkins. We thank the anonymous referee for helpful comments that improved this work.
MES is supported by the National Science Foundation Graduate Research Fellowship under grant number DGE 1256260. MM acknowledges support from NSF grant AST1312997. MW acknowledges support from NSF grants AST1313045 and AST1412999. EO acknowledges support from NSF grant AST1313006.

\facilities{MMT, Bok}

\end{document}